\documentclass[%
reprint,
showpacs,
nofootinbib,
 amsmath,amssymb,
prl
]{revtex4-1}

\usepackage{graphicx}
\usepackage{dcolumn}
\usepackage{bm}
\usepackage{color}
\newcommand{\vect}[1]{\mathrm{\textit{\textbf{#1}}}}

\renewcommand{\apj}{ApJ}
\newcommand{\mnras}{MNRAS}
\newcommand{\apjl}{ApJL}

\renewcommand{\nat}{Nature}
\newcommand{\aap}{A \& A}

\renewcommand{\prd}{Phys. Rev. D}

\newcommand{\Msun}{\mathrm{M}_{\odot}}

\begin{document}

\title{Black Hole Disks in Galactic Nuclei}

\author{\'Akos Sz\"olgy\'en}
 \email[Email: ]{szolgyen@caesar.elte.hu}
 \affiliation{Institute of Physics, E{\"o}tv{\"o}s University, P{\'a}zm{\'a}ny P.s., Budapest, 1117, Hungary}

\author{Bence Kocsis}
 \affiliation{Institute of Physics, E{\"o}tv{\"o}s University, P{\'a}zm{\'a}ny P.s., Budapest, 1117, Hungary}

\date{September 4, 2018}

\begin{abstract}
Gravitational torques among objects orbiting a supermassive black hole drive the rapid reorientation of orbital planes in nuclear star clusters (NSCs), a process known as vector resonant relaxation. In this Letter, we determine the statistical equilibrium of systems with a distribution of masses, semimajor axes, and eccentricities. We average the interaction over the apsidal precession time and construct a Monte Carlo Markov chain method to sample the microcanonical ensemble of the NSC. We examine the case of NSCs formed by 16 episodes of star formation or globular cluster infall. We find that the massive stars and stellar mass black holes form a warped disk, while low mass stars resemble a spherical distribution with a possible net rotation. This explains the origin of the clockwise disk in the Galactic center and predicts a population of black holes (BHs) embedded within this structure. The rate of mergers among massive stars, tidal disruption events of massive stars by BHs, and BH-BH mergers are highly increased in such disks. The first two may explain the origin of the observed G1 and G2 clouds, the latter may be important for gravitational wave detections with LIGO and VIRGO. More generally, black holes are expected to settle in disks in all dense spherical stellar systems assembled by mergers of smaller systems including globular clusters. 
\end{abstract}


\maketitle


\paragraph{Introduction.---}\hspace{-12pt} Many nearby galaxies show evidence of a supermassive black hole (SMBH) in their centers surrounded by a dense population of stars within a few parsec, a nuclear star cluster (NSC) \citep{Genzel2010,Kormendy2013}. At the Milky Way's center, the old low-mass stars are observed to be spherically distributed while young massive stars form a disk within the NSC \citep{Bartko2009,Lu2009,Yelda2014}. This structure is also apparent in other NSCs \citep{Seth2008}. 

The origin of the observed distribution is controversial. Some studies suggested that young stars are still found in the disk in which they were born, while the old stars' orbits have been randomized by the gravitational interaction \citep{Hopman2006}. However, since the timescale on which orbits reorient is shorter than the age of the massive young stars in the disk, the observed distribution of orbital planes of young massive stars must also resemble a statistical equilibrium \citep{Kocsis2011,Perets2018}. We examine the equilibrium anisotropy attained during this process, called vector resonant relaxation (VRR), in which massive objects are expected to settle in lower inclination orbits \citep{Rauch1996}. 

VRR governs the distribution of orbital planes in a NSC, represented by the angular momentum vector directions, $\hat{\bm{L}}_i$ \citep{Rauch1996}. It is driven by the gravitational torques among objects which orbit around the SMBH on eccentric orbits with periods of $1-10^4$ yr, and where the ellipses execute apsidal precession within their planes ($10^4$--$10^5$ yr) due to the dominant spherical component of the NSC's gravitational field. VRR causes $\hat{\bm{L}}_i$ to reorient within $10^5-10^7$ yr during which the distribution of eccentricity and semimajor axis, $e$ and $a$ vary little \citep{Kocsis2011}. Thus, relaxation is fundamentally different in the 6 phase space dimensions. The 2D subspace of mean anomaly and argument of pericenter is covered rapidly through phase mixing, the 2D space of $\hat{\bm{L}}$ directions reach an internal thermodynamic equilibrium due to VRR for fixed $e$ and $a$ \citep{Roupas2017}. Then as $e$ and $a$ diffuse slowly, the 2D distribution of $\hat{\bm{L}}$ evolves through a series of VRR equilibria. 

In this Letter, we determine the equilibrium distributions of orbital planes for NSCs with a wide distribution of mass, semimajor axis, and eccentricity. We orbit average the Hamiltonian on the apsidal precession timescale and construct a Monte Carlo Markov chain to generate the VRR equilbrium. We consider cases in which the cluster is assembled through a series of episodes in which stars are formed in disks within the NSC or they are deposited there by infalling globular clusters \citep{Tremaine1975,Milosavljevic2001,Antonini2012,Antonini2013,Gnedin2014,Antonini2014,Antonini2015,ArcaSedda2015,ArcaSedda2017}.

Besides explaining the observed NSCs, the statistical equilibria of VRR may have applications in high energy astrophysics and relativity. If the equilibrium configurations of massive objects are disks in NSCs, then stellar black holes (BHs) must also reside in these stellar disks. These BHs may be observed with x-ray instruments if they traverse gas clouds or form binaries \citep{Bartos2013}. They may also regulate gas accretion in active galactic nuclei (AGN) and grow into intermediate mass black holes (IMBHs) \citep{Kocsis2011b}. If they exist, disks of BHs cannot be neglected in precision tests of general relativity using stellar orbits \citep{Merritt2010,Meyer2012}. Furthermore, BHs merge  in disks more frequently and produce gravitational waves detectable by LIGO/VIRGO \citep{Kocsis2012,Bartos2017,Stone2017}. They also produce extreme mass ratio inspirals (EMRIs) detectable by LISA \citep{Gair2004,AmaroSeoane2007,Kocsis2011b,Babak2017}. 

The dynamics of VRR also shows tantalizing connections with systems in condensed matter physics, namely the Maier Saupe model of liquid crystals, vortex crystals, and the $N$-vector model of spin systems \citep{Kocsis2011,Kocsis2015}. The connection with liquid crystals may be understood as follows. On timescales longer than the orbit and apsidal precession, the orbits cover two-dimensional annuli. The pairwise time-averaged Newtonian Hamiltonian among these orbits is identical to the Coulomb interaction among axisymmetric molecules at the leading quadrupole order. In both cases, the objects align at low temperatures, form a nearly isotropic distribution at high temperatures, and undergo a first order phase transition \citep{Roupas2017,Takacs2018}. Higher multipoles dominate for low mutual inclination orbits, for which VRR resembles vortex crystals. The nearly coplanar limit yields the $N$-vector model of spin systems. VRR is an example of a nonadditive long-range interacting system which exhibits ensemble inequivalence \citep{Roupas2017}. VRR also allows negative absolute temperatures \citep{Kocsis2011,Roupas2017,Takacs2018}. Thus, VRR may have multidisciplinary applications.

\paragraph{Hamiltonian.---\label{Hamiltonian}} \hspace{-12pt} We utilize \textsc{n-ring} \citep{Kocsis2015} to calculate the effective Hamiltonian of VRR. In this framework, the Newtonian $N$-body Hamiltonian is averaged over orbits and the apsidal precession time for each pair of stars $i$ and $j$. The orbit-averaged Hamiltonian is
\begin{align}
\label{eq1}
H_{i j} &= \Bigg \langle - \frac{G m_i m_j}{|\vect{r}_i(t) - \vect{r}_j(t')|} \Bigg \rangle _{t,t'} \nonumber\\ &= - \frac{1}{P_{i} P_{j}} \oint d \vect{r}_i \oint d \vect{r}_j \frac{G m_i m_j}{v_i v_j |\vect{r}_i - \vect{r}_j|},
\end{align}
where $G$ is the gravitational constant, $m_i$ and $m_j$ are the masses, $\vect{r}_i$ and $\vect{r}_j$ are the position vectors of the gravitating object $i$ and $j$ which move along the precessing Keplerian trajectories around the central SMBH of mass $M$, $P_i = 2\pi (a_i^3 / GM)^{1/2}$ is the orbital period, $v_i = \sqrt{GM(2/|\bm{r}_i| - 1/a_i)}$ is the orbital speed. Averaging over apsidal precession time leads to the pairwise Hamiltonian between two solid bodies
\begin{equation}
\label{eq3}
H_{ij}= - G \int\limits_{S_i} d^2 r \int\limits_{S_j} d^2 r' \frac{ \sigma_i(r) \sigma_j(r')}{|\bm{r} - \bm{r'}|}\,,
\end{equation}
where $S_i$ is the 2D annulus covered by the orbit of the $i$th body in the plane perpendicular to $\hat{\bm{L}}_i$ between its peri- and apoapsis $r_{p,i}=(1-e_i)a_i$, $r_{a,i}=(1+e_i)a_i$, and the surface density is axisymmetric around $\hat{\bm{L}}_i$ with 
\begin{equation}
\label{eq2}
\sigma_i(r) = \frac{m_i}{2\pi^2 a_i \sqrt{(r_{a,i} - r)(r - r_{p,i})}} \qquad (r_{p,i}<r<r_{a,i})\,.
\end{equation}
Similar definitions apply for the $j$th body. Finally, the Hamiltonian is expanded in multipoles
\begin{equation}
\label{eq:H}
H = - \sum_{i,j}^{i < j} \sum_{\ell = 2}^{\infty} \mathcal{J}_{ij\ell} P_{\ell}(\hat{\vect{L}}_i \cdot \hat{\vect{L}}_j) ,
\end{equation}
where $\mathcal{J}_{ij\ell}$ are given by \textsc{n-ring} \citep{Kocsis2015}. These are constant coupling coefficients between orbital planes of the $i$th and the $j$th object corresponding to the $\ell$th Legendre polynomial, which depend on the masses, semimajor axes, and eccentricities. The dynamical variables of VRR are the angular momentum vector directions, $\hat{\bm{L}}_i$. Since $m_i$, $a_i$, and $e_i$ are drawn from distribution functions which do not change during VRR, $\mathcal{J}_{ij\ell}$ is a quenched random matrix \cite{Wu1982,Selinger2004}. We truncate the Hamiltonian at $\ell_{\max}=50$. The time evolution using the orbit-averaged VRR Hamiltonian \eqref{eq:H} has been verified against direct $N$-body methods using statistical measures \citep{Rauch1996,Eilon2009,Kocsis2015}.

\paragraph{Initial conditions.---} \hspace{-12pt} In all simulations we generate 16 disks with $N$ objects each, where $N=128$ or $512$, respectively in two sets of simulations, which represent 16 distinct star formation episodes or globular cluster infall events. We do not consider a possible additional primordially isotropic stellar distribution. In each disk the parameters of the objects, $(m_i,e_i,a_i)$ are drawn randomly from independent distributions proportional to $(m^{-2},e,a^0)$, respectively between limiting values with $m_{\max}/m_{\min}=100$, $e_{\min}=0$, $e_{\max}=0.3$, and $a_{\max}/a_{\min}=100$. A uniform $a$ distribution and $f(e)\propto e$ corresponds to a 3D number density in physical space $n(r)\propto r^{-2}$, which is expected for the equilibrium distributions of massive objects in NSCs \citep{Bachall1977}, and $e_{\max}$ is chosen to match their observed value \citep{Yelda2014}. The mass distribution approximately resembles the observed mass distribution of young stars \citep{Bartko2010}. To generate the initial disks, the angular momentum vectors are drawn from a uniform polar cap distribution around direction $\hat{\bm{n}}_{{\rm disk},K}$ in the region $\hat{\bm{L}}_i\cdot \hat{\bm{n}}_{{\rm disk},K}\geq 0.994$ for each $K\in\{1,\dots,16\}$ \citep{Bartko2010}. Here $\hat{\bm{n}}_{{\rm disk},K}$ are the symmetry axes of the 16 disks, which are drawn from an isotropic distribution (The VRR equilibrium is independent of the SMBH mass, which is expected to be reached for $M\lesssim 10^7\Msun$ \cite{Kocsis2011}.). We generate 100 such initial conditions.  

\paragraph{MCMC simulations.--- \label{MCMC}} \hspace{-15pt} We examine the microcanonical ensemble of VRR, assuming an isolated system of black holes and stars with fixed orbit-averaged total energy~\eqref{eq:H} and fixed total angular momentum $\bm{L}_{\rm tot}=\sum_i \bm{L}_i$. To do so, we created a Monte Carlo Markov chain, \textsc{n-ring-mcmc}, as follows. In each step the code selects random pairs of orbits from the ensemble and proposes to rotate their angular momenta $\bm{L}_i$ and $\bm{L}_j$ around their common total angular momentum $\bm{L}_i+\bm{L}_j$ by a random angle chosen uniformly between $[0,2\pi]$. Note that $\bm{L}_{\rm tot}$ of the system is conserved exactly by construction. Their $H_{ij}$ pairwise energy is also conserved, but the total energy of the system is not due to interactions with the rest of the orbits. We keep track of the cumulative change of the total energy relative to the initial value. The proposal is accepted if it is within a tolerance $\Delta E=\pm0.001\, G m_{\min}^2/a_{\min}$. The MCMC model calculates the interaction energies of Eq.~\eqref{eq:H}  using \textsc{n-ring} \cite{Kocsis2015}. These steps are repeated until the distribution of angular momentum vectors reach statistical equilibrium. Since the objects fill up the fixed total energy and angular momentum manifold in phase space, they sample the microcanonical ensemble. This applies in the region where external torques by sources surrounding the cluster are neglegible.

Since VRR describes a nonadditive system \citep{Campa2014,Roupas2017}, we generate two independent sets of MCMC simulations with different assumptions. In set (I), we generate $100$ separate NSCs, each of which are assembled by $16$ disks in distinct star formation or globular cluster infall episodes, as discussed above. Each disk contains $512$ objects. We add the disks one by one and allow the system to relax between each episode to find its statistical equilibrium configuration. In set (II), we generate another $100$ independent NSCs composed initially of $16$ disks, each with $128$ objects. Here, we start the simulation with all disks present at the same time. We find no qualitative differences between the two cases (I) and (II).

\begin{figure}[t]
\includegraphics[width=1\columnwidth,height=180pt]{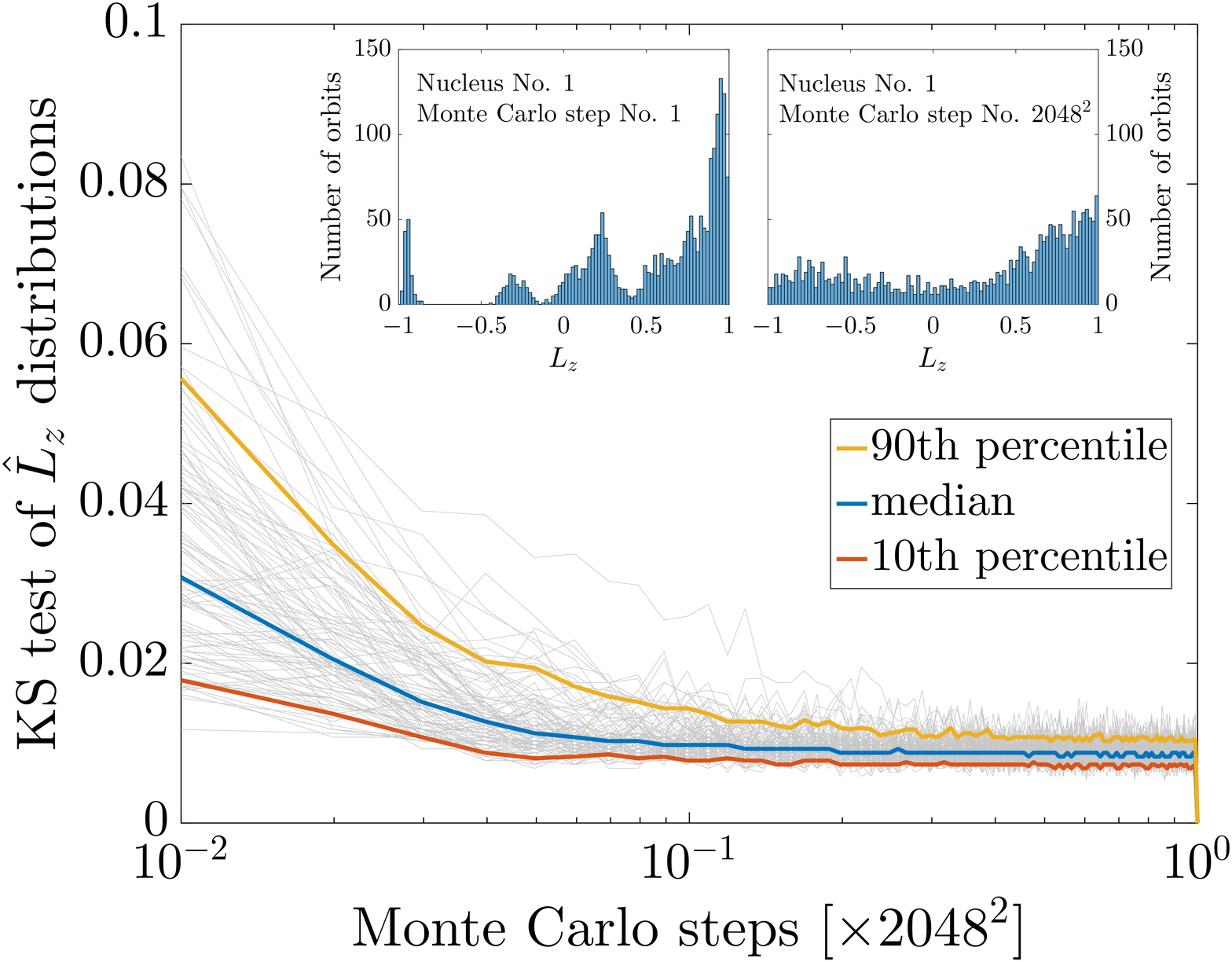}
\caption{\label{ConvTest} Convergence of the $\hat{L}_z$ distribution as a function of Monte Carlo steps. Faded curves show the Kolmogorov--Smirnov test of the $\hat{L}_z$ distributions for each simulation relative to the final distribution. Red, blue, and yellow curves show the 10th, 50th, and 90th percentile for each step. Insets show one example of the initial (left) and final (right) distribution. } \vspace{-15pt}
\end{figure}

\paragraph{Convergence.---} \hspace{-12pt} We tested the convergence of the $\hat{L}_z = \hat{\bm{L}}\cdot\hat{\bm{L}}_{\rm tot}$ distribution as a function of Monte Carlo steps by evaluating the Kolmogorov\--Smirnov test with respect to the final distribution. Figure \ref{ConvTest} shows the results for simulation set (II). Each gray curve represents a separate realization of the NSC, the red, blue, and yellow, curves show the $10$th, $50$th (median), and $90$th percentile of the results for each MCMC  step, respectively, among all $100$ simulations. The figure shows that the distributions typically converge well within the simulated number of steps. The insets show the initial and final $\hat{L}_z$ histogram for one example NSC in the simulation. 

\begin{figure*}
\includegraphics[width=0.45\textwidth]{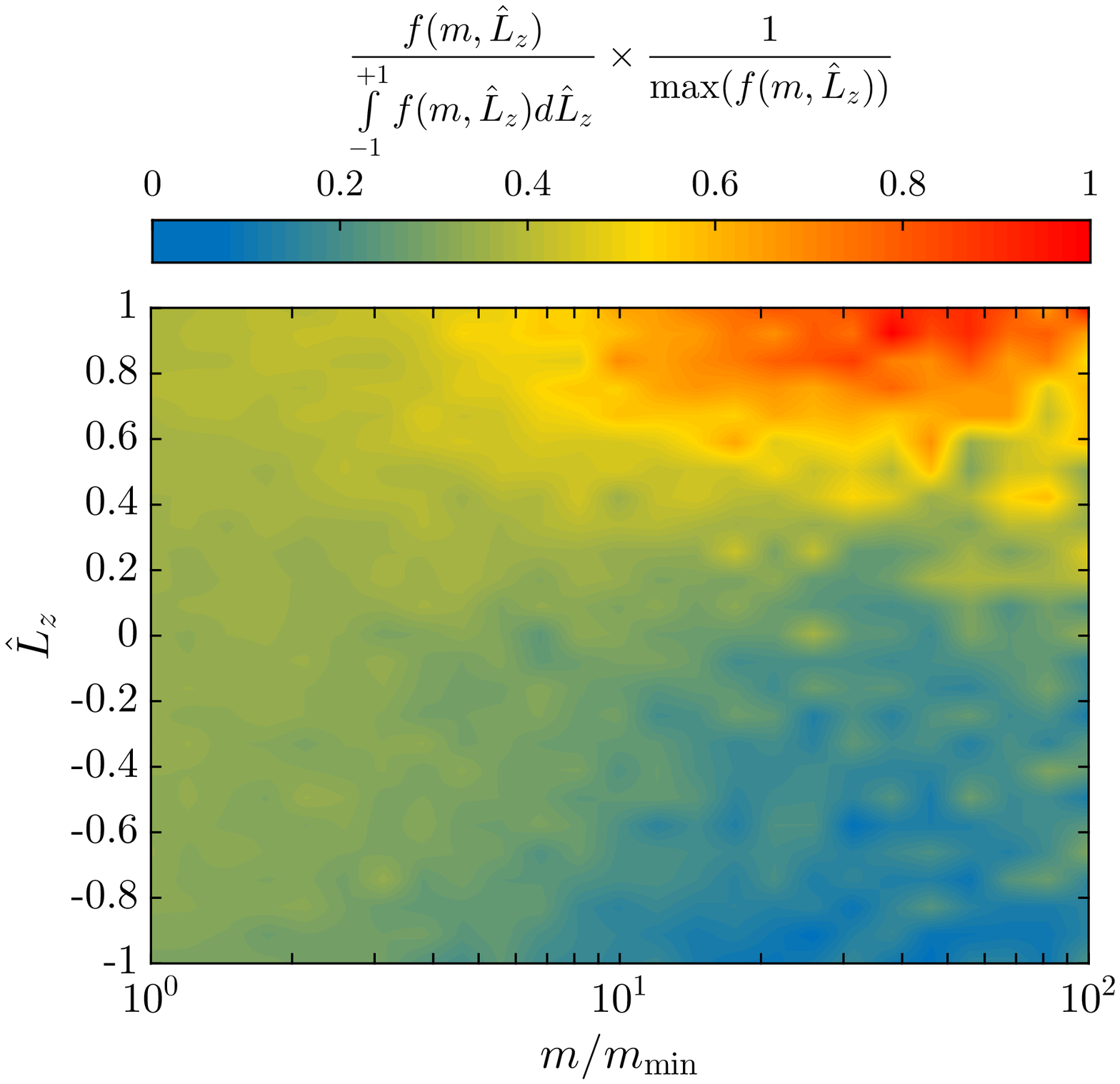}
\hspace{0.05\textwidth}
\includegraphics[width=0.45\textwidth]{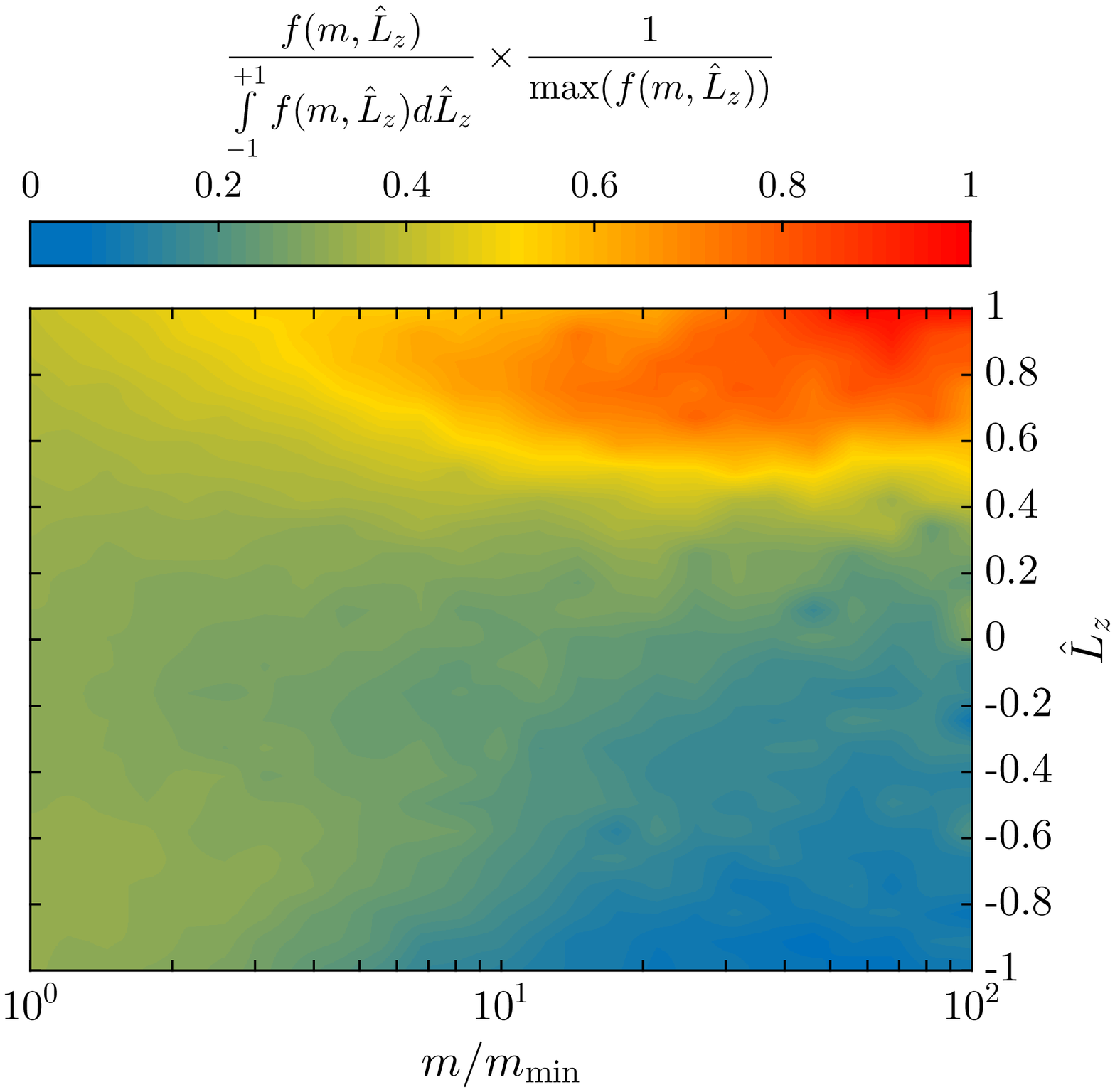}
\caption{\label{MassSeg}
Distribution of orbital planes ($\hat{L}_z = \hat{\bm{L}}\cdot\hat{\bm{L}}_{\rm tot}$) of objects orbiting around an SMBH as a function of object mass. The relaxed distribution, $f(m,\hat{L}_z)$ (see normalization in legend), is shown after 16 episodes of star formation or globular cluster infall events each delivering 128 objects (left panel) and 512 objects (right panel), respectively. The distribution of angular momentum vectors is isotropic for low masses and clustered around the total angular momentum of the cluster for high masses.}
\end{figure*}

\paragraph{Anisotropic mass segregation.--- \label{Anisotropic Mass Segregation}} \hspace{-15pt} We examine the relaxed distributions with respect to the masses of objects. Figure \ \ref{MassSeg} shows the distribution of $\hat{L}_z$ for objects in different mass bins for the two sets of simulations (I) and (II). In both cases low mass objects relax to an approximately isotropic distribution, however, higher mass objects display a strong anisotropy. Beyond a critical mass of around $5 m_{\min}$, the level of anisotropy increases with mass, implying a mass segregation in $\hat{\bm{L}}_z$ space. This resembles a thick disk of massive objects in physical space, where the disk thickness decreases with mass. A disk represents the nematic phase of liquid crystals \cite{Roupas2017,Takacs2018}.

\begin{figure*}
\includegraphics[width=0.315\textwidth]{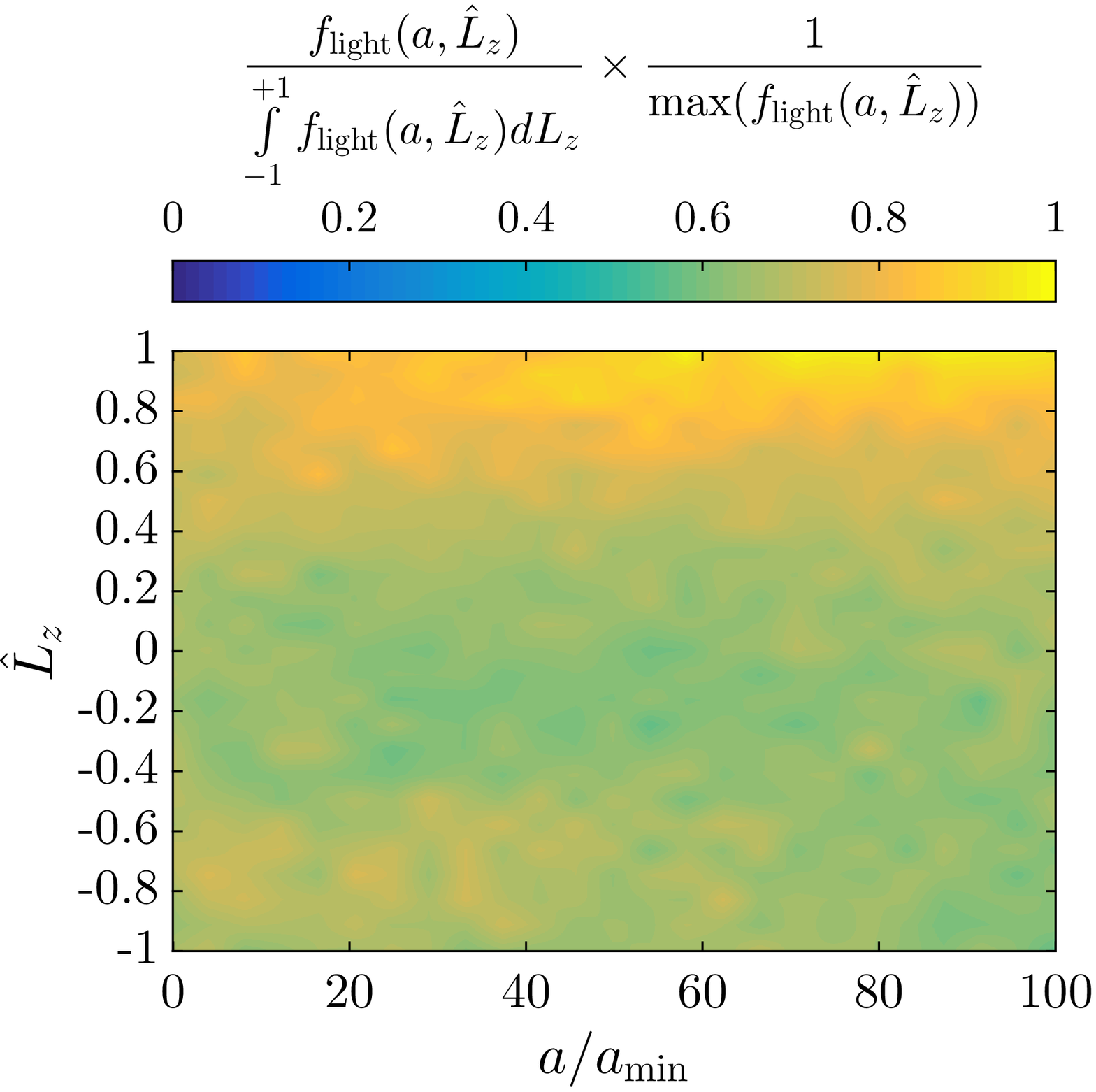} 
\hspace{0.025\textwidth}
\includegraphics[width=0.27\textwidth]{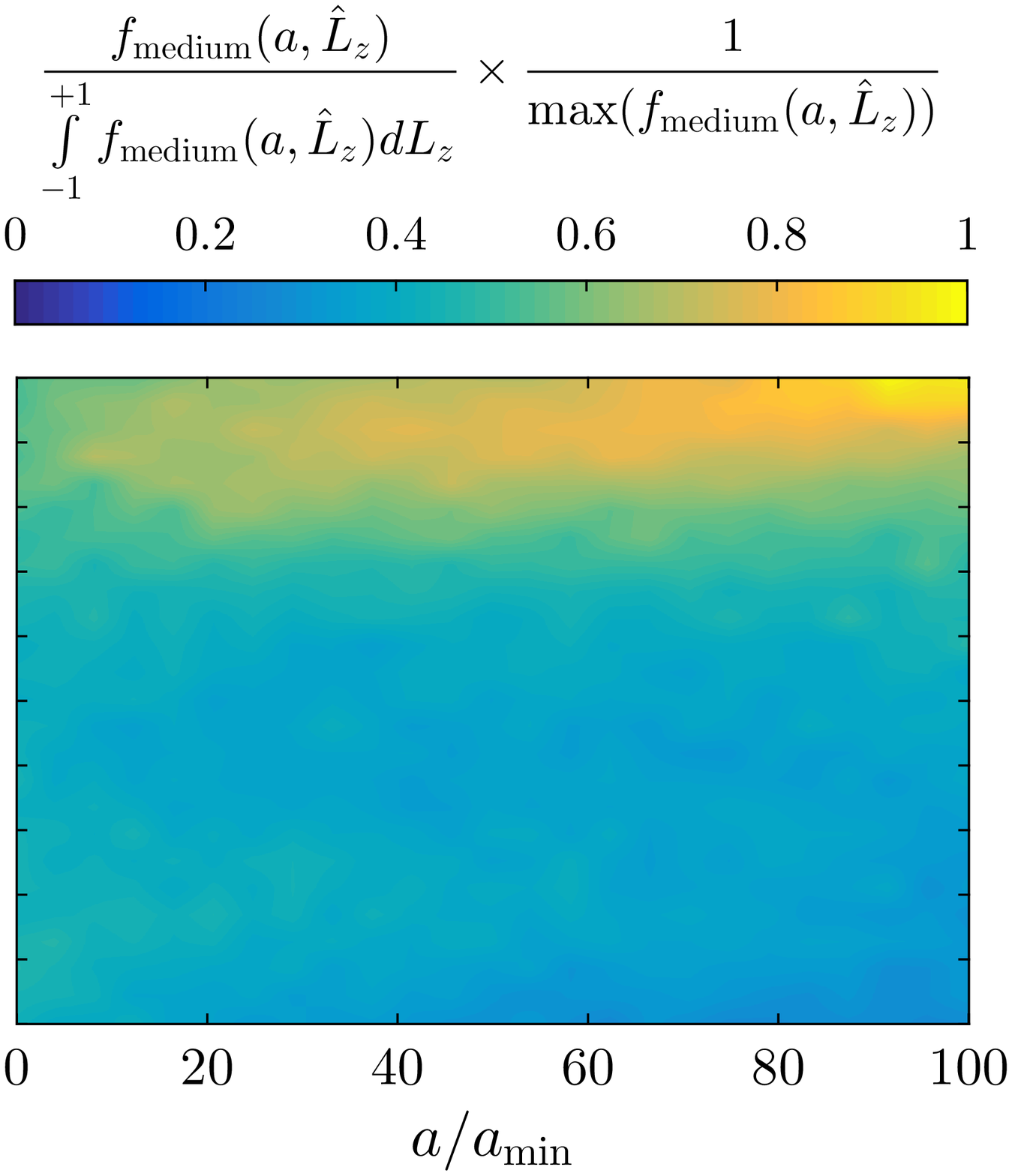}
\hspace{0.03\textwidth}
\includegraphics[width=0.31\textwidth]{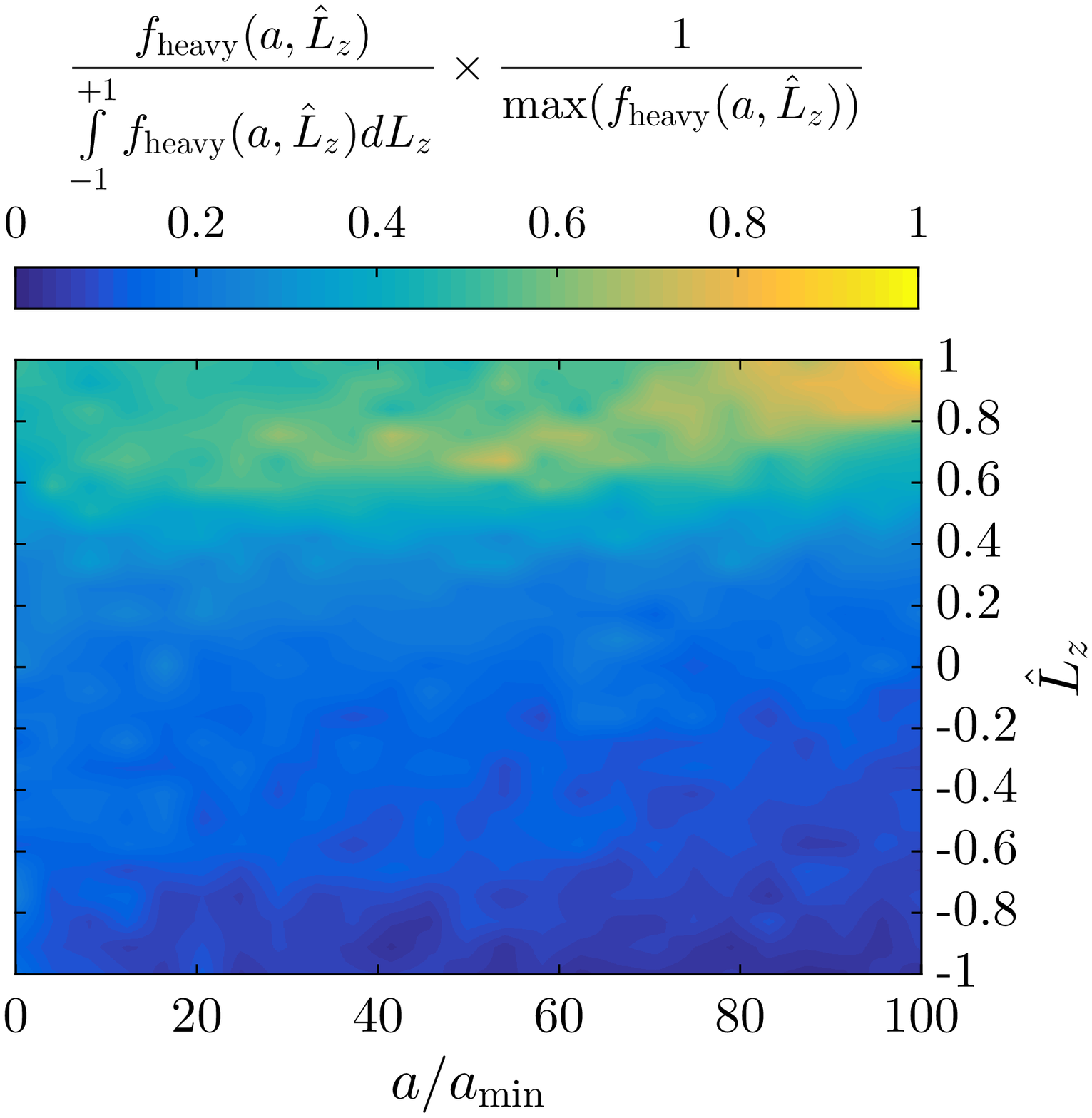}
\caption{\label{Warp} Distribution of orbital planes of objects as in Fig.~\ref{MassSeg} but as a function of the semimajor axis ($a$) in $3$ different mass bins, respectively: 
$1 m_{\min} - 2 m_{\min}$ (left panel), $2 m_{\min} - 16 m_{\min}$ (middle panel), $16 m_{\min} - 100 m_{\min}$ (right panel).
The result for low mass objects corresponds to a spherical distribution with a net rotation, the intermediate mass objects resemble a thick disk and a spherical population, and the heaviest objects resemble a warped disk which is most prominent in the outer region.
}
\end{figure*}

Do the disk thickness and inclination depend on semimajor axis? We group the objects by mass into three bins between $1\Msun-2\Msun$,  $2\Msun-16\Msun$, $16\Msun-100\Msun$, which, respectively, correspond to old main sequence (MS) stars and neutron stars (NSs), \textit{B} stars and low mass BHs, and \textit{O} stars and heavy BHs. Figure~\ref{Warp} shows that low mass objects (left panel) are mostly isotropic, but they show an excess angular momentum at larger semimajor axis. This implies a spherical distribution with a net rotation. The intermediate mass and high mass objects (middle panel) show a strong clustering of $\hat{\bm{L}}$ around $\hat{\bm{L}}_{\rm tot}$ (i.e., $\hat{L}_z\sim 1$). There is a systematic change in the mean of $\hat{L}_z$ with $a$, which in physical space represents a warped disk. In the highest mass bin (right panel) the disk is most apparent in the outer region and shows a strong warp.

\paragraph{Conclusion.---} \hspace{-12pt} In this letter, we examined the statistical equilbrium of orbital planes of gravitating objects around SMBHs in galactic nuclei. We considered the formation of the NSC by 16 episodes of star formation or infalling globular cluster each depositing a distribution of objects in axisymmetric disk configurations with a random symmetry axis. We constructed Monte Carlo Markov chains to generate the microcanonial ensemble of VRR with fixed total energy and angular momentum. The results showed that the distribution of low-mass objects becomes spherical, but the velocity distribution retains a net rotation, particularly in the middle and outer regions. The distribution of intermediate mass objects is anisotropic, showing a thick disk and a spherical distribution. The massive objects settle into a warped disk. The disk is thinner for higher masses, and is most prominent in the outer regions of the NSC. The disk of massive objects is a result of resonant dynamical friction \citep{Rauch1996}. The final outcome is expected to depend on the initial conditions mainly through the mass function and the number of infall episodes, $K$, where the disk is more prominent for smaller values of $K$. (For higher $K$, the magnitude of total VRR energy per star is smaller, which results in less anisotropy \cite{Roupas2017,Takacs2018}. Further, if the mass fraction in the initial disks is negligible compared to a primordial spherical NSC, a disk is not expected to form.)

These findings are consistent with observations of the Milky Way's center. These show a spherical cluster of low mass stars with a net rotation \citep{Feldmeier2014}, a warped clockwise disk of massive \textit{O}-type stars \citep{Bartko2009,Lu2009,Yelda2014}, and an anisotropic angular momentum distribution of \textit{B} stars \citep{Yelda2014}. A detailed investigation of the Milky Way's center, using time-dependent simulations will be presented elsewhere \citep{Kocsis2018}. 

Our results have implications for the distributions of NSs and BHs orbiting around SMBHs in NSCs which may be identified in x-ray and radio observations \citep{Bower2014,Spitler2014,Hailey2018}. Our results show that NSs are expected to be isotropically distributed, while the distribution of BH orbits resembles a warped thick disk. The symmetry axis of the BH disk is parallel with that of the massive stars, and the disk thickness depends only on the BH mass. BHs with masses comparable to that of the observed \textit{O} stars settle in the same distribution as \textit{O} stars. Thus, in the Milky Way's center, stellar mass BHs are not isotropically distributed, but they are expected to reside in the observed clockwise disk of WR and \textit{O}-type stars assuming that VRR equilibrium has been reached \citep{Bartko2009,Yelda2014,Kocsis2011}. Intermediate mass black holes (IMBHs), if they exist, are also expected to reside in this structure. A study of VRR exploring other distributions of mass, semimajor axis, and eccentricity is underway. Further, note that the likelihood of massive stellar collisions, tidal disruptions of massive stars by stellar BHs, and stellar BH-BH mergers are strongly enhanced if they reside in disks. Thus, the observed G1 and G2 gas clouds in this disk \citep{Gillessen2012,Phifer2013,Pfuhl2015,Witzel2017} may be remnants of collisions of two massive stars or that with a stellar BH. The likelihood of BHs crossing and illuminating these clouds \citep{Bartos2013} is also enhanced. BH disks in NSCs may provide abundant sources of GWs for LIGO/VIRGO and LISA \citep{OLeary2009,Kocsis2012,Leigh2018}. 

More generally, since this process operates in all dense stellar systems with a spherical potential \citep{Rauch1996}, BH disks may also be expected to be common in globular clusters \cite{meiron2018}. This depends on the level of anisotropy during star formation. Massive objects retain the original anistropy and settle in a disk due to resonant dynamical friction, while the distribution of low mass objects becomes spherical. Our results suggest that globular clusters formed by mergers \citep{Fujii2012} inevitably form BH disks. However, process driven by two-body encounters such as core collapse, runaway collisions \citep{PortegiesZwart2002,PortegiesZwart2004} and the BH retention fraction \citep{Morscher2013,Morscher2015} may be strongly modified in such configurations.

\acknowledgments{
We thank Jean-Baptiste Fouvry and Scott Tremaine for helpful comments. This work was supported by the \'UNKP-17-3 New National Excellence Program of the Hungarian Ministry of Human Capacities and received funding from the European Research Council (ERC) under the European Union's Horizon 2020 research and innovation programme under grant agreement No 638435 (GalNUC) and by the Hungarian National Research, Development, and Innovation Office grant NKFIH KH-125675. The calculations were carried out on the NIIF HPC cluster at the University of Debrecen, Hungary.}

\bibliography{aszolgyen_2018}

\end{document}